\documentclass[aps,prc,notitlepage,%
longbibliography,twocolumn]{revtex4-1}

\usepackage{natbib}
\usepackage{graphicx}
\usepackage{amssymb}
\usepackage{amsmath}
\usepackage{ifthen}
\usepackage{braket}
\usepackage{xcolor}
\usepackage{bm}
\usepackage{bbm}
\usepackage{comment}
\usepackage{siunitx}
\usepackage{float}
\usepackage{dcolumn}
\newcolumntype{d}[1]{D{.}{.}{#1}}
\usepackage{subfigure}

\newcommand{\dd}{\mathrm{d}}

\usepackage{xcolor}
\usepackage{bbm}

\definecolor{light}{gray}{0.90}
\definecolor{darker}{gray}{0.50}
\definecolor{dark}{gray}{0.30}

\begin{document}

\title{Proton Radius: A Puzzle or a Solution!?}

\author{Ulrich D.~Jentschura}

\affiliation{Department of Physics and LAMOR, Missouri University of Science and
Technology, Rolla, Missouri 65409, USA}

\affiliation{MTA–DE Particle Physics Research Group, P.O. Box 51, 
H-–4001 Debrecen, Hungary}

\affiliation{MTA Atomki, P.O. Box 51, H–4001 Debrecen, Hungary}

\begin{abstract}
The proton radius puzzle
is known as the discrepancy of the proton radius,
obtained from muonic hydrogen spectroscopy 
(obtained as being roughly equal to 
$0.84$\,fm), and the proton
radius obtained from (ordinary) hydrogen spectroscopy 
where a number of measurements involving highly excited
states have traditionally favored a value of about $0.88$\,fm. 
Recently, a number of measurements of
hydrogen transitions by the Munich (Garching)
groups (notably, several hyperfine-resolved 
sublevels of the $2S$--$4P$) 
and by the group at the University of Toronto
($2S$--$2P_{1/2}$) have led to transition frequency data
consistent with the smaller proton radius of about $0.84$\,fm. 
A recent measurement of the $2S$--$8D$ transition
by a group at Colorado State University 
leads to a proton radius of about $0.86$\,fm,
in between the two aforementioned results.
The current situation points to a possible,
purely experimental, resolution of the proton radius puzzle. 
However, a closer look at the situation reveals
that the situation may be somewhat less clear, raising the question
of whether or not the proton radius puzzle has been conclusively solved, and
opening up interesting experimental possiblities at TRIUMF/ARIEL.
\end{abstract}

\maketitle

\section{Introduction}

We present a brief account of the current status 
of measurements of the proton radius, 
based on recent hydrogen spectroscopy~\cite{BeEtAl1997,ScEtAl1999,JeKoLBMoTa2005,%
PoEtAl2010,Je2015muonic,BeEtAl2017,BeEtAl2019,BrEtAl2022}
and scattering experiments~\cite{BeEtAl2010,XiEtAl2019}.
Also, we point out that a current 
discrepancy of the deuteron radius, 
based on deuterium spectroscopy and scattering
experiments, merits further investigation~\cite{PoEtAl2016}.
Attention is drawn to the fact that in 1969,
coindicidentally on the same day on which mankind
set foot on the moon (21st of July, 1969), two Letters were
published~\cite{CaEtAl1969prl1,CaEtAl1969prl2},
which report on an observation
of a larger slope of the Sachs form factor of the 
proton, when measured using electron versus 
muon scattering, consistent with a (roughly) 2\% 
difference in the proton radii derived from either 
scattering method.
The current situation opens interesting experimental possibilities
for the MUSE experiment~\cite{MUSE,KoEtAl2014,Ko2022priv} at the Paul--Scherrer Institute 
(PSI) and for a potential muon option at 
the Advanced Rare Isotope Laboratory (ARIEL) at TRIUMF (Vancouver).

%
%
%
\section{Brookhaven Experiment}

It might be useful to recall that scattering 
experiments on proton targets, using 
either electron or proton projectiles, 
have a considerable history. One example dates 
from about five decades ago~\cite{CaEtAl1969prl1,CaEtAl1969prl2}.
It was observed that muon-proton ($\mu$-$p$) versus 
electron-proton ($e$-$p$) scattering data 
for the Sachs form factor taken in the range
$0.15 \, ({\rm GeV}/c)^2 < q^2 < 0.85 \, ({\rm GeV}/c)^2$
differ from each other by a relative factor 
$N = 0.960 \pm 0.006$.
It is stated, in the right text column
on p.~154 of Ref.~\cite{CaEtAl1969prl1}, that the 
``4\% suppression of the form-factor ratio represents
an 8\% difference in the cross section 
since $\dd \sigma/\dd q^2 \propto G^2$''.

One should note that the authors of 
Ref.~\cite{CaEtAl1969prl2} base their analysis on a number of 
assumptions. Let us consider the 
Rosenbluth formula [see Eq.~(1) of Ref.~\cite{CaEtAl1969prl1}],
\begin{multline}
\label{cross_section}
\left. \frac{\dd \sigma}{\dd q^2} \right|_{\mu,e} =
\left. \frac{\dd \sigma}{\dd q^2} \right|_{\rm NS}
\frac{1}{ \cot^2(\theta/2)} \,
\\
\times \left[ 2 \tau G_M(q^2) + 
\frac{ G_E^2(q^2) + \tau \, G_M(q^2)}{1 + \tau} \,
\cot^2(\theta/2) \right] \,,
\end{multline}
where $\tau = -q^2/(4 m_p^2) = \vec q^2/(4 m_p^2)$
(for purely space-like momentum transfer),
$\theta$ is the scattering angle,
and $\left. \dd \sigma/\dd q^2 \right|_{\rm NS}$
is the differential cross section for spinless 
particles. The proton radius is inferred
in Ref.~\cite{CaEtAl1969prl2} as follows
[see also Eq.~(11) of Ref.~\cite{Je2011radii}],
\begin{equation}
r_p = \sqrt{ \langle r^2 \rangle_p } \,,
\qquad
\langle r^2 \rangle_p = r_p^2 =
6 \hbar^2 \left.
\frac{\partial G_E(q^2)}{\partial q^2} \right|_{q^2 = 0}
\end{equation}
The experiment~\cite{CaEtAl1969prl1,CaEtAl1969prl2}
has been carried out 
in a region of small scattering angle $\theta$,
where the second (slope) term in square
brackets in Eq.~\eqref{cross_section} 
dominates the first
(intercept), and the electric and magnetic 
form factors cannot be determined separately.
The authors of of Refs.~\cite{CaEtAl1969prl2}
base their analysis on the 
assumption that $G_E = G_M/\mu \equiv G$,
where $\mu \approx 1$ is the ratio of 
the magnetic and electric form factors of the 
proton. This assumption is rather well
justified and is in fact confirmed by
more recent experiments~\cite{JoEtAl2000prl}.

In the abstract of Ref.~\cite{CaEtAl1969prl2}, it 
is stated that the ``apparent disagreement 
[of the $\mu$-$p$ versus $e$-$p$ data]
can possibly be accounted for by a 
combination of systematic normalization errors''.
This statement highlights a possible
problem in the experiment~\cite{CaEtAl1969prl1,CaEtAl1969prl2}:
Namely, it is experimentally difficult to 
accurately normalize the flux of incoming particles,
which could in principle account for the
tentative muon-proton nonuniversality observed in 
Refs.~\cite{CaEtAl1969prl1,CaEtAl1969prl2}.
Yet, as evident from the rather detailed analysis 
present in the two Letters~\cite{CaEtAl1969prl1,CaEtAl1969prl2},
it is evident that considerable efforts were 
made by the authors of 
Refs.~\cite{CaEtAl1969prl1,CaEtAl1969prl2} to avoid such 
normalization errors.
In fact, in Ref.~\cite{CaEtAl1969prl1}, the scattering data
were checked for the possible presence of inelastic and
multiple-scattering contributions,
and also, the validity of the functional form~\eqref{cross_section}
was verified against experimental data,
for both muon as well as anti-muon projectiles.
Let us also record an observation: The 
differential cross section is roughly 
proportional to the square of the Sachs electric $G_E$ 
form factor (under the assumptions made above),
and the slope of the Sachs form factor
is again proportional to the square of the 
proton radius.
So, the observed (roughly) 8\% difference of the 
$\mu$-$p$ scattering cross sections~\cite{CaEtAl1969prl1,CaEtAl1969prl2}
translates into a 4\% difference of the 
form factor, which translates into a 2\% difference
of the proton radii derived from 
$\mu$-$p$ versus $e$-$p$ scattering.
Tentatively, one could argue that,
if the trend seen in the 1969 experiment
were to be exclusively due to a normalization
error, the relative systematic error would have to
be on the order of 8\% for the cross section, which is quite large.

Coincidentally, this is exactly the difference
(both sign and magnitude agree)
between the proton radii derived from the 
most recent $2S$--$8D$ measurement on
atomic hydrogen~\cite{BrEtAl2022} and the 
result from muonic hydrogen spectroscopy~\cite{PoEtAl2010}.
This situation gives a very pronounced motivation
for the MuSE experiment~\cite{MUSE,KoEtAl2014,Ko2022priv}, 
and also for adding a possible
muon option to TRIUMF/ARIEL.

\begin{figure}[t]
\begin{center}
\includegraphics[width=0.99\linewidth]{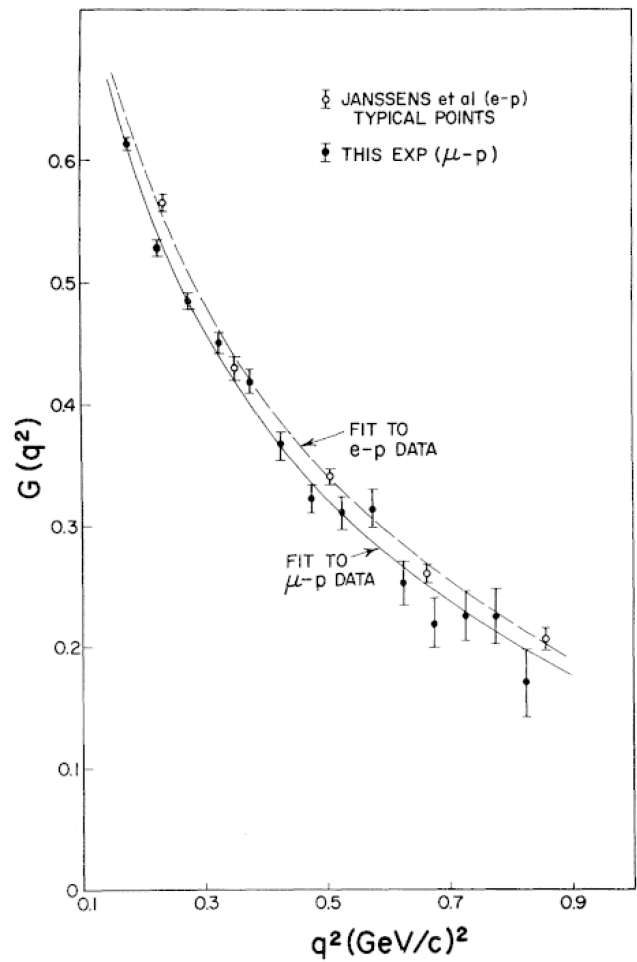}
\end{center}
\caption{\label{fig1}
[ Reproduced with permission. ]
Cross section data from Refs.~\cite{CaEtAl1969prl1,CaEtAl1969prl2}.}
\end{figure}

\section{Brief Review of Hydrogen Experiments}

In contrast to scattering experiments, the 
proton radius is derived from spectroscopic 
experiments via the finite-size energy shift,
which, in leading order and applying the 
nonrelativistic approximation, reads as 
\begin{equation}
\label{DeltaE}
\Delta E = \frac{2}{3} \, \frac{( Z \alpha )^4 \, \mu c^2}{\pi \, n^3} \,
\delta_{\ell 0} \,
\left( \frac{\mu  \, c \, r_p}{\hbar} \right)^2 \,,
\end{equation}
where $Z$ is the nuclear charge number
of the hydrogenlike ion, $\alpha$ is the fine-structure constant, 
$\mu$ is the reduced mass,
$c$ is the speed of light,
and $\ell$ is the orbital angular momentum.
The Kronecker $\delta$ implies that the 
energy correction is nonvanishing only for $S$ states
with $\ell = 0$.
The relation~\eqref{DeltaE} is only the leading approximation
to the finite-size effect. Higher-order terms 
involve convoluted moments of the nuclear-charge distribution,
as discussed in Chap.~5 of Ref.~\cite{JeAd2022book}
and in great detail in Ref.~\cite{Fr1979}.

Let us discuss Eq.~\eqref{DeltaE}. We have $Z=1$ for both hydrogen and muonic
hydrogen. The reduced mass $\mu$ is roughly 200 times larger for muonic bound
systems as compared to ordinary hydrogen.  The finite-size effect is
proportional to $\mu^3$, and thus, muonic hydrogen energy levels constitute a
very sensitive probe of the proton radius.

In chronological order, let us discuss some recent 
results for the proton radius from 
spectroscopic~\cite{BeEtAl1997,ScEtAl1999,JeKoLBMoTa2005,%
PoEtAl2010,Je2015muonic,BeEtAl2017,BeEtAl2019,BrEtAl2022},
as well as scattering, experiments~\cite{BeEtAl2010,XiEtAl2019}.
We first recall that a 
least-squares fit of all experimental data 
available up to 2005 for both hydrogen as well 
as deuterium led to the results $r_p = 0.8750(68) \, {\rm fm} 
\approx 0.88 \, {\rm fm}$
and $r_d = 2.1394(28) \, {\rm fm}$, 
based on a total of twenty-three measured transitions,
including transitions to highly excited states~\cite{JeKoLBMoTa2005}.

In 2010, the muonic hydrogen $2S$--$2P_{1/2}$ Lamb shift measurement at
PSI (Ref.~\cite{PoEtAl2010}) led to a proton radius 
of $r_p = 0.84184(67)\,{\rm fm} \approx \, 0.84\,{\rm fm}$.
The result was in disagreement with the results of an 
electron scattering experiment carried 
out in Mainz (in 2010, see Ref.~\cite{BeEtAl2010}), 
which led to the result of 
$r_p = 0.879(8)\,{\rm fm} \approx \, 0.88\,{\rm fm}$,
consistent with the value obtained from 
atomic hydrogen spectroscopys~\cite{JeKoLBMoTa2005}.

Since 2010, the discrepancy 
between the proton radius obtained from
hydrogen spectroscopy
($r_p \approx 0.88 \, {\rm fm}$) and from
muonic hydrogen spectroscopy
($r_p \approx 0.84 \, {\rm fm}$) 
has been termed the proton radius puzzle.
In 2017, a measurement of the $2S$--$4P$
transition in atomic hydrogen (Ref.~\cite{BeEtAl2017}, 
taking into account various fine-structure and hyperfine sublevels)
led to a result of $r_p = 0.8335(95)\,{\rm fm}$,
consistent with the smaller proton radius,
otherwise obtained from muonic hydrogen spectroscopy.
In the analysis of the experiment~\cite{BeEtAl2017}, 
emphasis is laid on so-called cross-damping terms,
which had been analyzed (for differential 
cross sections) in Ref.~\cite{JeMo2002}.

A recent hydrogen measurement (in 2018, by the French group)
of the $1S$--$3S$ transition~\cite{FlEtAl2018}
is consistent with the larger proton radius,
with a result of $r_p = 0.877(13)\,{\rm fm} \approx 0.88 \, {\rm fm}$.
In Ref.~\cite{Je2015muonic}, it had been pointed out that 
the experimental approach taken by the 
Paris group should be largely independent of cross-damping terms.

The PRad Scattering Experiment, in 2019,
obtained a value of $r_p = 0.831(14)\,{\rm fm}$,
as reported in Ref.~\cite{XiEtAl2019},
based on scattering data.
In 2019, a spectroscopic measurement in 
Toronto, of the $2S$--$2P_{1/2}$ Lamb shift in atomic hdrogen,
resulted in a proton radius of 
$r_p = 0.833(10)\,{\rm fm}$~\cite{BeEtAl2019},
consistent with the smaller value of the proton radius.
One might point that 
the $2P_{1/2}$--$2P_{3/2}$ fine-structure 
is nearly independent of the proton radius and can be 
calculated to very high precision~\cite{JePa1996};
its measurement would constitute an important 
consistency check for the smallness of the ``Canadian protons''
measured in Ref.~\cite{XiEtAl2019}.
(Of course, the Einstein equivalence principle (EEP) implies that
the outcome of an experiment should be independent
of where and when in the Universe it is performed.
Assuming that protons are indistinguishable, 
identical elementary particles, one might be tempted to include the 
tongue-in-the-cheek remark that
any conceivable size difference in 
``French versus Canadian and German Protons''
would violate the EEP.)

In Ref.~\cite{Je2015muonic}, it had been pointed out that
the so-called cross-damping terms, 
which had played an important role in the 
analysis of the $2S$--$4P$ experiment~\cite{BeEtAl2017}, 
should be suppressed in $2S$--$nD$ transitions
in atomic hydrogen [see the discussion surrounding 
Eqs.~(28)---(31) of Ref.~\cite{Je2015muonic}].
It had also been pointed out in Ref.~\cite{Je2015muonic}
that cross-damping terms are ``not likely to shift the
experimental results reported in Refs.~\cite{BeEtAl1997}
and~\cite{ScEtAl1999} on a level commensurate with the proton radius puzzle
energy shift.'' We recall that Refs.~\cite{BeEtAl1997} and~\cite{ScEtAl1999}
report on measurements of $2S$--$nD$ transitions with 
$n=8$ and $n=12$. In consequence, it is of significance
that a very recent (2022) measurement of the $2S_{1/2}$--$8D_{5/2}$ 
transition in atomic hydrogen (see Ref.~\cite{BrEtAl2022})
has led to a proton radius of 
$r_p = 0.8584(51)\,{\rm fm} \approx 0.86 \, {\rm fm}$,
which lies in between the previously 
accepted value of $r_p \approx 0.88 \,{\rm fm}$
from atomic hydrogen spectroscopy~\cite{JeKoLBMoTa2005}
and the value of $r_p \approx 0.84 \,{\rm fm}$ 
derived from muonic hydrogen spectroscopy~\cite{PoEtAl2010}.

The proton radius puzzle is exacerbated
by the fact that the raw data for the 
Sachs form factor derived from the 
2010 Mainz experiment~\cite{BeEtAl2010}
and the 2019 PRad data~\cite{XiEtAl2019}
are discrepant~\cite{Be2022priv}
(see also~\cite{Je2022talk}).
Clearly, further investigations are indicated.

%
%
\section{Conclusions}

In view of interesting possibilities with muon physics,
one might ask if one 
should add a muonic beam to TRIUMF/ARIEL.
One idea for low-cost muon sources can 
be mentioned, being based
on an 8\,GeV proton beam on a current-carrying target 
followed by a lithium lens and a quadrupole
decay channel, where the decaying ions 
generate muons~\cite{BaMo2011}.

It is highly unlikely that 
muonic hydrogen theory, and Lamb shift theory, 
could provide explanations for the proton radius 
puzzle, since they are well under control.
From the experimental side, the 
situation regarding the 
proton radius may be less clear than commonly thought.
Electron versus muon scattering experiments could shed light.
A confirmation of the 1969 experiment~\cite{CaEtAl1969prl1,CaEtAl1969prl2},
which could point to a small electron-muon nonuniversality,
would be very desirable.
A comparison of electron and muon scattering on the same apparatus
would shed light on the issue~\cite{MUSE}.
Finally, let us remember that an even more striking 
$7 \sigma$ disagreement persists for the deuteron radius,
when comparing the results from electronic and 
muonic bound systems~\cite{PoEtAl2016}.

%
\section*{Acknowledgments}

Insightful conversations with J. C. Bernauer
are gratefully acknowledged.
This research was supported by NSF grant PHY--2110294.


\begin{thebibliography}{10}

\bibitem{BeEtAl1997}
B. de~Beauvoir, F. Nez, L. Julien, B. Cagnac, F. Biraben, D. Touahri, L.
  Hilico, O. Acef, A. Clairon, and J.~J. Zondy, Phys. Rev. Lett. {\bf 78},  440
   (1997).

\bibitem{ScEtAl1999}
C. Schwob, L. Jozefowski, B. de~Beauvoir, L. Hilico, F. Nez, L. Julien, F.
  Biraben, O. Acef, J.~J. Zondy, and A. Clairon, Phys. Rev. Lett. {\bf 82},
  4960  (1999), [Erratum Phys. Rev. {\bf 86}, 4193 (2001)].

\bibitem{JeKoLBMoTa2005}
U.~D. Jentschura, S. Kotochigova, E.-O. Le~Bigot, P.~J. Mohr, and B.~N. Taylor,
  Phys. Rev. Lett. {\bf 95},  163003  (2005).

\bibitem{PoEtAl2010}
R. Pohl {\it et~al.}, Nature (London) {\bf 466},  213  (2010).

\bibitem{Je2015muonic}
U.~D. Jentschura, Phys. Rev. A {\bf 92},  012123  (2015).

\bibitem{BeEtAl2017}
A. Beyer, L. Maisenbacher, A. Matveev, R. Pohl, K. Khabarova, A. Grinin, T.
  Lamour, D.~C. Yosta, T.~W. H\"{a}nsch, N. Kolachevsky, and T. Udem, Science
  {\bf 358},  79  (2017).

\bibitem{BeEtAl2019}
N. Bezginov, T. Valdez, M. Horbatsch, A. Marsman, A.~C. Vutha, and E.~A.
  Hessels, Science {\bf 365},  1007  (2019).

\bibitem{BrEtAl2022}
A.~D. Brandt, S.~F. Cooper, C. Rasor, Z. Burkley, A. Matveev, and D.~C. Yost,
  Phys. Rev. Lett. {\bf 128},  023001  (2022).

\bibitem{BeEtAl2010}
J.~C. Bernauer, P. Achenbach, C. Ayerbe~Gayoso, R. B\"{o}hm, D. Bosnar, L.
  Debenjak, M.~O. Distler, L. Doria, A. Esser, H. Fonvieille, J.~M. Friedrich,
  J. Friedrich, M. G\'{o}mez Rodr\'{i}guez de~la Paz, M. Makek, H. Merkel,
  D.~G. Middleton, U. M\"{u}ller, L. Nungesser, J. Pochodzalla, M. Potokar, S.
  S\'{a}nchez~Majos, B.~S. Schlimme, S. Sirca, T. Walcher, and M. Weinriefer,
  Phys. Rev. Lett. {\bf 105},  242001  (2010).

\bibitem{XiEtAl2019}
W. Xiong {\it et~al.}, Nature (London) {\bf 575},  147–150  (2019).

\bibitem{PoEtAl2016}
R. Pohl {\it et~al.}, Science {\bf 353},  669  (2016).

\bibitem{CaEtAl1969prl1}
L. Camilleri, J.~H. Christenson, M. Kramer, L.~M. Lederman, Y. Nagashima, and
  T. Yamanouchi, Phys. Rev. Lett. {\bf 23},  149  (1969).

\bibitem{CaEtAl1969prl2}
L. Camilleri, J.~H. Christenson, M. Kramer, L.~M. Lederman, Y. Nagashima, and
  T. Yamanouchi, Phys. Rev. Lett. {\bf 23},  153  (1969).

\bibitem{MUSE}
see https://www.psi.ch/en/muse/experiment.

\bibitem{KoEtAl2014}
\relax{M. Kohl {\em et al.} [MUSE Collaboration]}, Eur. Phys. J. Web of
  Conferences {\bf 66},  06010  (2014).

\bibitem{Ko2022priv}
M. Kohl, private communication (2022).

\bibitem{Je2011radii}
U.~D. Jentschura, Eur. Phys. J. D {\bf 61},  7  (2011).

\bibitem{JoEtAl2000prl}
M.~K. Jones {\it et~al.}, Phys. Rev. Lett. {\bf 84},  1398  (2000).

\bibitem{JeAd2022book}
U.~D. Jentschura and G.~S. Adkins, {\em \relax{Quantum Electrodynamics: Atoms,
  Lasers and Gravity}} (World Scientific, Singapore, 2022).

\bibitem{Fr1979}
J.~L. Friar, Ann. Phys. (N.Y.) {\bf 122},  151  (1979).

\bibitem{JeMo2002}
U.~D. Jentschura and P.~J. Mohr, Can. J. Phys. {\bf 80},  633  (2002).

\bibitem{FlEtAl2018}
H. Fleurbaey, S. Galtier, S. Thomas, M. Bonnaud, L. Julien, F. Biraben, F. Nez,
  M. Abgrall, and J. Gu\'{e}na, Phys. Rev. Lett. {\bf 120},  183001  (2018).

\bibitem{JePa1996}
U. Jentschura and K. Pachucki, Phys. Rev. A {\bf 54},  1853  (1996).

\bibitem{Be2022priv}
J.~C. Bernauer, private communication (2022).

\bibitem{Je2022talk}
U.~D. Jentschura, Talk given at the workshop ``New Scientific Opportunities and
  the TRIUMF ARIEL e-linac'', TRIUMF, Vancouver, Canada, 26th of May, 2022.

\bibitem{BaMo2011}
V. Balbekov and N. Mokhov, {\em Low--Budget Muon Source}, PAC2001, TPAH144,
  Particle Accelerator Conference, FNAL, Batavia, Illinois, 2001.

\end{thebibliography}
\end{document}